\documentstyle[12pt,graphicx]{article}

\title{ Using  polarized maser
 to detect high-frequency relic gravitational waves  }

\author{\small  M L Tong$^1$, Y Zhang$^1$\thanks{Electronic address:
yzh@ustc.edu.cn},
                      \ and   F Y Li$^2$\\
  \small $^1$ Astrophysics Center,
      University of Science and Technology of China,
      Hefei, Anhui, 230026,  China\\
  \small $^2$ Dept. of Physics,
      Chongqing University, Chongqing, 400044, China }
 \date{}

\def\be{\begin{equation}}
\def\ee{\end{equation}}
\def\ba{\begin{eqnarray}}
\def\ea{\end{eqnarray}}

\def\x{\mathbf{x}}

\def\d{\mathrm{d}}

\begin{document}
\maketitle

\baselineskip=18truept

\normalfont

\begin{center}
\Large  Abstract
\end{center}

\begin{quote}

A  GHz maser beam with Gaussian-type distribution passing through a
homogenous static magnetic field can be used to detect gravitational
waves (GWs) with the same frequency. The presence of GWs will
perturb the electromagnetic (EM) fields, giving rise to perturbed
photon fluxes (PPFs). After being reflected by a fractal membrane,
the perturbed photons suffer little decay and can be measured by a
microwave receiver.
This idea has been explored to certain extent as
a  method for very high frequency gravitational waves.
In this paper,
we examine and develop this method more extensively,
and confront the possible detection with the predicted signal
of relic gravitational waves (RGWs).
A maser beam with high linear polarization  is used
to reduce  the background photon fluxes (BPFs) in the detecting
direction  as the main noise. As a key factor of applicability of this
method, we  give a preliminary estimation of  the sensitivity of a sample
detector limited by thermal noise using currently common technology.
 The minimal detectable amplitude of GWs
is found to be $h_{\rm{min}}\sim10^{-30}$. Comparing with the known
spectrum of the RGWs in the accelerating universe for $\beta=-1.9$, there is still
roughly a gap of $4\sim 5$ orders. However, possible improvements on the
detector can further narrow down the gap and make it a feasible
method to detect high frequency RGWs.

\end{quote}

\newpage

\begin{center}

{\bf I. INTRODUCTION}
\end{center}

Although there has been some indirect evidence of GWs radiation from
the binary pulsar B1913+16 \cite{taylor}, so far GWs have not been
directly detected  yet. Currently, a number of detectors have been
running or under construction. These detectors use various methods
including: (1) the conventional method of cryogenic resonant bar
\cite{bar} aiming at a frequency around $10^3$ Hz; (2) the method of
ground-based laser interferometers, such as LIGO \cite{Abramovici},
VIRGO \cite{Bradaschia}, GEO \cite{GEO}, TAMA \cite{tama}, and AIGO
\cite{chunnong},
 applying for a frequency range  $10\sim10^3$ Hz,
 and the space-based laser
interferometers  LISA \cite{Jafry} under planning for a lower
frequency range $(10^{-3}, 10^{-1})$ Hz; (3) the detections of
cosmic microwave background radiation (CMB) polarization of
``magnetic'' type, or  the temperature-``electric''
cross-correlation \cite{baskaran}, which would also give direct
evidence of GWs \cite{Bpol} for very low frequencies around
$(10^{-18}, 10^{-16})$ Hz. Besides, there have also been attempts,
based on various techniques, to detect GWs of very high frequencies
from MHz to GHz, such as the waveguide detector to measure the
change of polarizations of EM waves \cite{cruise2, Tong},  the
two coupled microwave cavities to measure small harmonic
displacements \cite{Bernard}, and laser interferometers \cite{Akutsu}, etc.

There is another method of detection for high frequency
gravitational waves  (HFGWs), which employs a a maser beam  passing
through a strong static magnetic field \cite{Li,li1,li4}, and uses a
microwave receiver in combination with a fractal membrane
\cite{Baker,li2}. The maser beam can be chosen to a free electron
maser  \cite{Cohen,Abramovich} with a great output power $\sim 2$
kW, whose frequency  $\sim 4.5$ GHz is the one that the fractal
membrane operates effectively \cite{Wen,Zhou,Hou}. In the presence
of GWs, the background EM fields will be perturbed, giving rise to
additional photon fluxes in various directions. In the previous
studies by Li et al \cite{li4,li2}, an ordinary maser beam is used,
in which case the BPFs always exist and tend to mix up with the
PPFs, forming a kind of noise. Moreover, in order to assess the
method as a potential way to detect GWs, the sensitivity of the
detection predicted by this method has to be estimated, and a
comparison with the predicted spectrum of RGWs
\cite{Grishchuk1}-\cite{Miao} is still needed.
 In this paper, we improve the
method by using a linearly polarized maser beam, so that  the BPFs
in the detecting direction can be suppressed effectively.

Generally speaking,  HFGWs  in GHz band are
not generated by usual astrophysical processes, such as explosions
of asymmetric supernovas, rotations of  binary stars around each
other, coalescing and merging of binary neutron stars or black
holes, and collapse of stars \cite{grishchuk5,Flanagan,Zhang3}.
There could be a thermal background of gravitational waves,
 which consists of gravitons in thermal equilibrium
\cite{Zeldovich,buonanno}.
But, it will be examined that,
if the inflationary expansion did occur  in the  early universe,
the possible thermal GWs will be negligibly small.
As is known, RGWs generated by
the inflation have a spectrum stretching over a whole range of
$(10^{-18}, 10 ^{11})$ Hz.
In particular, it has a considerable amount of power around
the very high frequency range $\sim 4.5$ GHz.
Thus it can serve as the main target of detection
using the maser beam.
We shall estimate qualitatively
the sensitivity of detection of a sample detector,
and make a
comparison with the known analytic spectrum of RGWs \cite{Zhang2, Zhang1,Miao},
which has not been made before.

The organization of the paper is as follows. In Section II,
 we describe a setup of maser-beam GW detection.
In Section III  we study  the PPFs
density caused by the incident GWs propagating along various directions,
and estimate the number of the perturbed photons per second passing
through a receiving surface.
Section IV is devoted to a
preliminary analysis of the sensitivity of the detector limited by
the thermal noise,
and to the discussion of the feasibility of
detecting the RGWs in the accelerating universe.
In Section V, a
summary is  given.
The Appendix gives a detailed calculation  of the
PPFs density generated by the incident GWs along the positive $z$-direction.

\begin{center}
{\bf II.  THE SETUP OF THE DETECTOR}

\end{center}

\begin{figure}
\centerline{\scalebox{1.0}[1.0]
{\includegraphics[width=6cm]{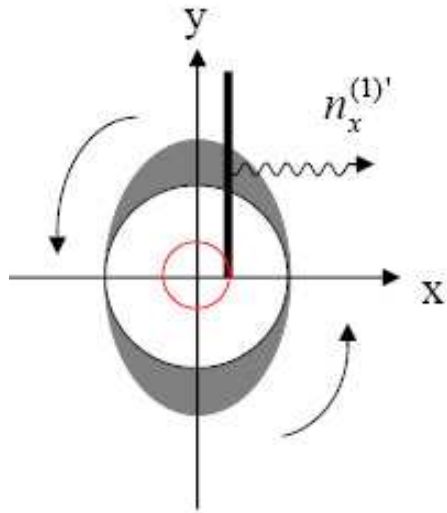}}} \caption{\label{sketch} A
sketch map of the circular perturbed photon flux. The inner circle
stands for the maser beam. The dark areas stand for the perturbed
photon flux in the circular direction, where the thicker region
means a larger flux. The membrane is placed near the Gaussian beam
and the flux is  reflected back into the positive $x$-direction.}
\end{figure}

The idea  of the maser beam gravitational wave detector is based on
the property that the maser beam in the presence of a  homogeneous
static magnetic field will be perturbed when GWs
pass by \cite{li1,li2}.
In particular, under the resonance  condition that the frequency
 of GWs   equals that of the maser beam ($\nu_g=\nu_e$),
additional PPFs will be generated and serve  as  a signal of GWs to
be detected. As can be seen later, the   magnitude of the PPFs  is
proportional to the amplitude of GWs and to the static magnetic
field as well. Fig.\ref{sketch1} shows the geometric configuration of the setup for the
detector.  The maser beam of frequency $\nu_e \sim4.5$ GHz travels  along
the positive $z$-direction and passes through the static magnetic field
 $\sim$ 3 Tesla  pointing to the positive $y$-direction.
 The PPF density (photons per unit area per unit time) in the $x$-direction,  $n_x^{(1)}$,  after being totally
 reflected by a fractal membrane \cite{Wen,Zhou,Hou},
 will keep its strength  constant within a distance
of about $1$ meter. The reflected PPF density $n_x^{(1)'}$
 will be received by a microwave receiver as a
signal  of GWs. As a typical pattern,  some of the  generated PPFs
will travel around the maser beam \cite{li1}, forming a circular
flux, which is shown in Fig.\ref{sketch}.

Although the maser beam is set to propagate along the
positive $z$-direction overwhelmingly,
there is always a leakage flux
 of photons along the radial direction (normal to the $z$-direction).
These leaking photons  will mix
up with the perturbed photons and form a noise for
detection.
Since we have chosen to detect the PPFs in the $x$-direction,
we should try to eliminate the  BPFs in the $x$-direction.
This  can be achieved by using a linearly
polarized  maser beam.
Maser beams in the GHz band have been generated under laboratory
conditions \cite{Cohen,Abramovich}.
In our setup of the detector,
we make use of a maser beam with transverse electric mode ($\mathrm{TEM}_{00}$),
whose strength has a Gaussian-type distribution \cite{Yariv}:
\begin{figure}
\centerline{\scalebox{1.0}[1.0]
{\includegraphics[width=10cm]{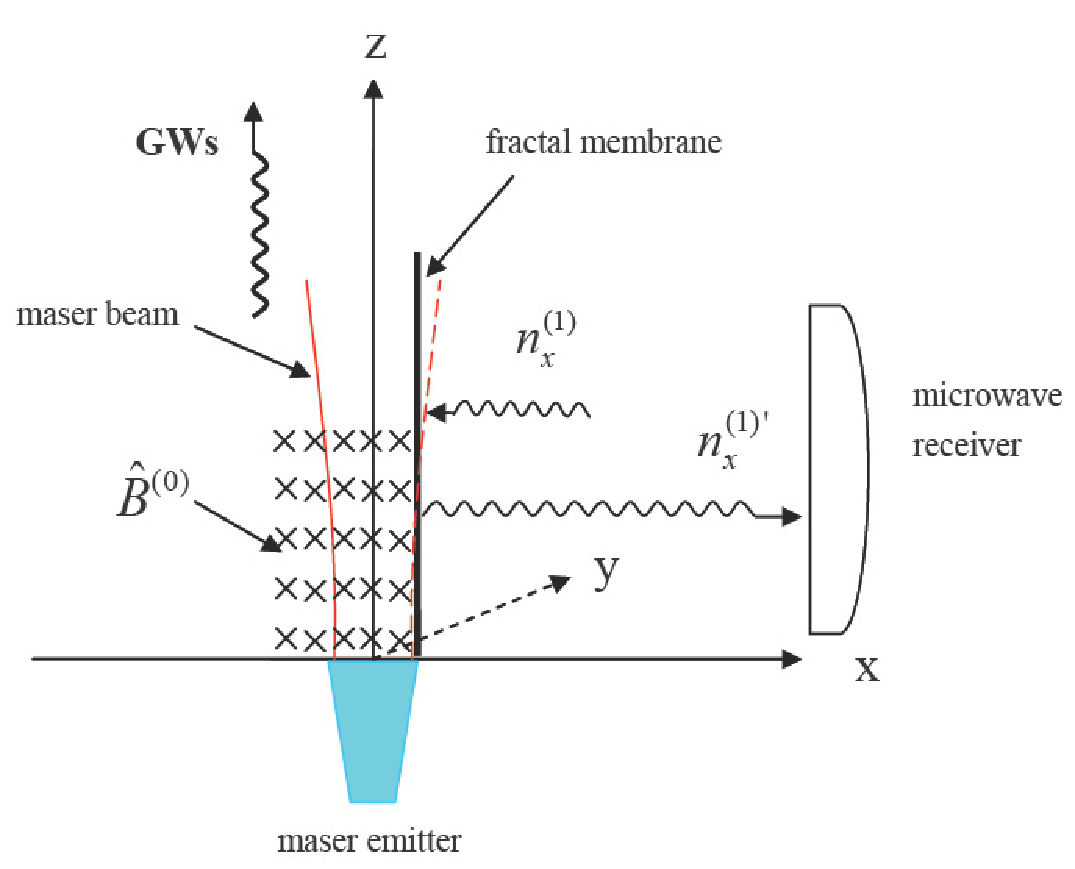}}} \caption{\label{sketch1}
The sketch map of the setup of detection. The maser beam is
propagating along the positive $z$-direction, and the static
magnetic field
 points to  the positive $y$-direction as represented
by the furcations.
 The envelope of the maser beam
is sketched as  the hyperbola. The fractal  membrane is placed in
the $y-z$ plane near the maser beam, facing to the right.
The portion of the PPFs  along the
negative $x$-direction is reflected by the fractal membrane.
The reflected PPF density $n_x^{(1)'}$ goes along the positive $x$-direction
and keep its strength constant within 1 meter. The microwave receiver is
placed on the right, facing to the left, and catching the outgoing
perturbed photons as the signal. }
\end{figure}
\be\label{psi}
\psi({\bf x},t)
=\frac{\psi_0}{\sqrt{1+(z/f)^2}} \exp(-\frac{r^2}{W^2})
\exp\left\{i[(k_ez-\omega_et)-\tan^{-1}{\frac{z}{f}}
+\frac{k_er^2}{2R}+\delta]\right\},
 \ee
where $\psi_0$ is the amplitude on the plane $z=0$,
$r=\sqrt{x^2+y^2}$ is the radial distance, $W =W_0[1+(z/f)^2]^{1/2}$
with $W_0$ being the radius of the beam on the plane $z=0$, $f=
W_0^2 k_e/2$, $k_e=2\pi/\lambda_e$ is the wavenumber,
$\omega_e$ is the angular frequency,
$R=z+f^2/z$ is the curvature radius of the
wave fronts of the beam at $z$,
and $\delta$ is an arbitrary phase factor.
Furthermore, the maser beam  is linearly polarized along the
$x$-direction, namely, the electric field in the maser beam is given by
\be \label{initial}
  \tilde{E}_x^{(0)}({\bf  x},t)=\psi({\bf x},t),\ \qquad
\tilde{E}_y^{(0)} = \tilde{E}_z^{(0)}=0,
\ee
where the tilde ``$\sim $'' and the superscript $``(0)$'' stand for the
time-dependent and the background EM fields, respectively.
Since
the maser beam emitted from the emitter is in the region $z\geq0$.
the static magnetic field is chosen to localize in the region $z\geq0$,
\ba \label{initial2}
&&\hat{B}^{(0)}=
 \left\{\begin{array}{cc}
\hat{B}_y^{(0)}&(0\leq z\leq l), \\
0&\ \ \qquad (z<0\ \ {\mathrm{or}} \ \ z> l),
\end{array}\right.
\ea
where the caret ``$\wedge$'' denotes the static magnetic field, and
$l$ is the dimension of the static magnetic field in
the $z$-direction.

In absence of GWs, the components of the average  BPF density
($m^{-2}s^{-1}$) are given as follows,
\ba\label{nx0}
 n_x^{(0)}&=&0,\\
 \label{ny0}
n_y^{(0)}& = & -\frac{1}{ \mu_0 \hbar\omega_e}
           \langle \tilde{E}_x^{(0)} \tilde{B}_z^{(0)}\rangle \nonumber \\
       &=& \frac{\psi_0^2k_ey}{2\mu_0 \hbar \omega_e^2[1+(z/f)^2]
           (z+f^2/z)}\exp{(-\frac{2\,r^2}{W^2})},\\
n_z^{(0)}&=&\frac{1}{\mu_0 \hbar\omega_e}
         \langle \tilde{E}_x^{(0)}\tilde{B}_y^{(0)}\rangle \nonumber   \\
 &=&\frac{\psi_0^2}{2\mu_0\hbar \omega_e^2[1+(z/f)^2]}\left[k_e+
 \frac{k_er^2(f^2-z^2)}{2(f^2+z^2)^2}-\frac{f}{f^2+z^2}\right]
 \exp{(-\frac{2\,r^2}{W^2})},
 \ea
where ``$\langle\rangle$''
means the average over a time scale much longer than  $1/\nu_e$.
Note that, since the maser has been chosen to be linearly polarized
in $x$-direction,
the $x$-component of the  BPF density,
$n_x^{(0)}$ as a noise,  is vanishing.
This feature is an advantage over that using an
unpolarized beam \cite{li2}.
Of course, in actual situation, the maser beam can not be polarized
completely. Then, Eq.(\ref{nx0}) is not valid strictly, and it always
exits the residual BPF density $n_x^{(0)}$.
Just like the components $n_y^{(0)} $ and  $n_z^{(0)}$,
$n_x^{(0)}$  will decay by a factor $e^{-\frac{2r^2}{W^2}}$,
moreover, the fractal membrane only reflects the PPF not the BPF \cite{li2}.
However, after reflected by the fractal membrane,
$n_x^{(1)}$ will not decay within  1 meter.
 Then, at a large radial distance $r$ from the beam,
 $n_x^{(0)}$ can be negligible compared
 with  $n_x^{(1)}$.
We will discuss  about this  problem in more details in Sec III.

\begin{center}
{\bf III. PPFs GENERATED BY GWs ALONG  VARIOUS DIRECTIONS}

\end{center}

\begin{center}
{ \bf A.   GWs along positive $z$-direction }
\end{center}

The situation we are interested in is  when GWs are present.
Consider a beam of circularly polarized
 monochromatic plane GWs propagating along
 the positive $z$-axis.
The metric can be written as
\be\label{metric}
 g_{\mu\nu}=\eta_{\mu\nu}+ h_{\mu\nu},
  \ee where
$\eta_{\mu\nu}$ is  Minkowsky metric,
and  $ h_{\mu\nu}$  stands for GWs with $| h_{\mu\nu}|\ll 1$.
In TT (transverse-traceless) gauge,
$ h_{\mu\nu}$ has only two
independent components: $h_{11}=-h_{22}\equiv h_\oplus$, and
$h_{12}=h_{21}\equiv h_\otimes$,
where
 \ba\label{circular}
&&h_\oplus=A_\oplus\exp{[i(k_gz-\omega_gt)]},\nonumber\\
&&h_\otimes=iA_\otimes\exp{[i(k_gz-\omega_gt)]}.
 \ea
In a curved spacetime,
the  Maxwell's equations in vacuum are \cite{mtw,weinberg}
\ba\label{Mexwell} &&(\sqrt{g}\,
g^{\mu\alpha}g^{\nu\beta}
               F_{\alpha\beta})_{,\, \nu}=0,\\ \label{Mexwell2}
&&F_{\mu\nu,\,\sigma}+F_{\nu\sigma,\,\mu}+F_{\sigma\mu,\,\nu}=0,
\ea
where $F_{\mu\nu}$ is the EM  field tensor,
$g\equiv-{\mathrm{det}}(g_{\mu\nu})$,
and the comma means the ordinary derivative.
Since the EM field will be perturbed by GWs,
we decompose the total EM tensor   into two parts:
 \be\label{emtensor}
F_{\mu\nu}=F_{\mu\nu}^{(0)}+F_{\mu\nu}^{(1)},
 \ee
where  $F_{\mu\nu}^{(0)}$ represents the background fields, and
$F_{\mu\nu}^{(1)}$  the perturbed fields caused by GWs.
Explicitly, \ba\label{fcovariant1}
 &&F^{(0)}_{\mu\nu}=\frac{1}{c}
 \left(\begin{array}{cccc}
0  & - \tilde{E}_x^{(0)}  & 0 & 0\\
\tilde{E}_x^{(0)}  & 0 & c\tilde{B}_z^{(0)}  &
           -c(\hat{B}_y^{(0)}+\tilde{B}_y^{(0)})\\
0  &-c\tilde{B}_z^{(0)}     & 0 & c\tilde{B}_x^{(0)}\\
0  &  c(\hat{B}_y^{(0)}+\tilde{B}_y^{(0)} )
        & -c\tilde{B}_x^{(0)}
        & 0\end{array}\right), \nonumber\\
 \label{fcovariant2}
&& F^{(1)}_{\mu\nu}=
 \frac{1}{c}\left(\begin{array}{cccc}
0 & - \tilde{E}_x^{(1)}
  & - \tilde{E}_y^{(1)}
  & - \tilde{E}_z^{(1)}\\
\tilde{E}_x^{(1)}  & 0
  & c\tilde{B}_z^{(1)}& -  c\tilde{B}_y^{(1)}\\
\tilde{E}_y^{(1)}
  & -c\tilde{B}_z^{(1)}  & 0 & c\tilde{B}_x^{(1)}\\
\tilde{E}_z^{(1)}
  & c\tilde{B}_y^{(1)}
  & -c\tilde{B}_x^{(1)} & 0\end{array}\right).
\ea
Since  $|h_{\mu\nu}|\ll 1$,
$F_{\mu\nu}^{(1)}$ is also small and
will be evaluated up to the first order of $|h_{\mu\nu}|$.
As will be seen,
each component $F_{\mu\nu}^{(1)}$ receives two parts of contributions:
one comes from the interaction between the static
magnetic field and the GWs, $\propto |h_{\mu\nu}| \hat { B}^{(0)}$,
the other comes from the interaction between the maser beam and the
GWs, $\propto |h_{\mu\nu}| \tilde { B}^{(0)}$ \cite{Codina,Logi,Li}.
In our designing of the detection,
the static magnetic field is
chosen to be so large that $\tilde{B}^{(0)}/\hat{B}^{(0)}
\sim10^{-5}$~\cite{li2}.
Therefore, in $F_{\mu\nu}^{(1)}$ we only
keep the contribution $\propto |h_{\mu\nu}| \hat { B}^{(0)}$.
 The maser beam just provides the resonance condition  to generate the
PPFs, i.e., the detector  only respond  to the GWs with  the same
frequency as the maser beam.
By solving the  Maxwell's equations
(\ref{Mexwell}), one obtains the perturbed EM fields and
the PPFs.
The detailed calculation of $n_x^{(1)}$ is given in Appendix.
The resulting  expressions of
 the PPFs density $n_x^{(1)}$ in all three regions
 are given by

 \

Region I $(z\leq0)$.
 \be \label{nx1region1}
n_x^{(1)}=0;
 \ee

  Region II $(0\leq z\leq l)$,
  \ba\label{nx1region2}
&&n_x^{(1)}= -\frac{A_\otimes
\hat{B}_y^{(0)}\psi_0y}{2\mu_0\hbar\omega_e[1+(z/f)^2]^{1/2}}
   \left\{\frac{k_g z}{2(z+f^2/z)}
   \sin{ \Phi }\right.\nonumber\\
&&\qquad\quad\left.  +\frac{z}{W_0^2[1+(z/f)^2]} \cos{ \Phi}
 +\frac{1}{2(z+f^2/z)}\sin{(k_gz)} \sin{ \left(k_gz+\Phi
\right) }\right.\nonumber\\
&&\qquad\quad\left.+\frac{1}{k_gW_0^2[1+(z/f)^2]}\sin{(k_gz)}
\cos{\left(k_gz +\Phi \right)}
    \right\}\exp{\left(-\frac{r^2}{W^2}\right)};
\ea

\begin{figure}
\centerline{\scalebox{1.0}[1.0]
{\includegraphics[width=12cm]{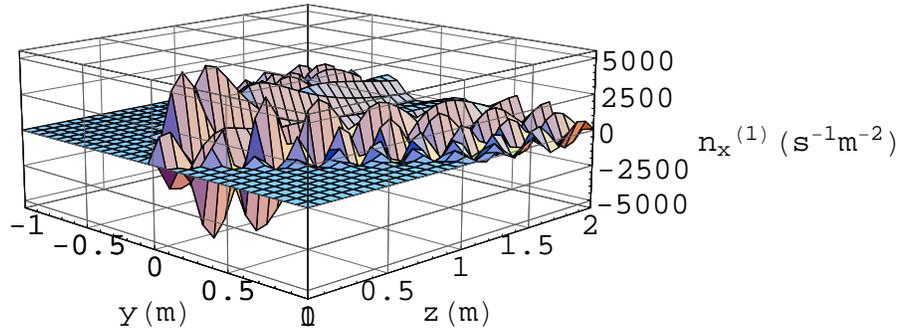}}}
\caption{\label{nxdependyz}
 $n_x^{(1)}$  as a function of $(y,z)$ on the plane  $x=0.05$ m.
  The parameters are taken  as $A_\otimes\sim10^{-30}$,
  $\psi_0=1.8\times 10^4$\,
V\,m$^{-1}$,\ $W_0=0.05 $\,m,
 \ $\hat{B}_y^{(0)}=3$\,T, and $l =0.4\,$m for demonstration.}
\end{figure}

\begin{figure}
\centerline{\scalebox{1.0}[1.0]
{\includegraphics[width=12cm]{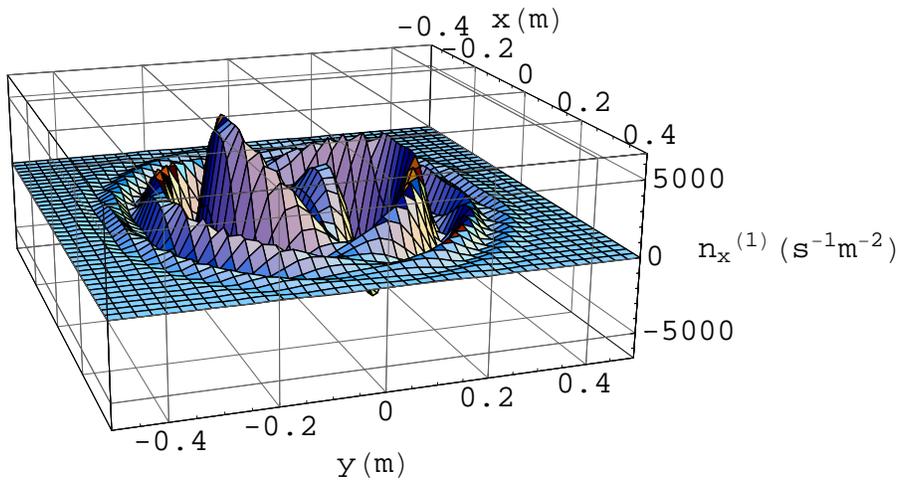}}}
\caption{\label{nxdependyz1}
 $n_x^{(1)}$  as a function of $(x,  y)$ on the plane  $z=0.4$ m.
  The parameters are taken the same as in Fig.\ref{nxdependyz}.}
\end{figure}

  Region III $(z\geq l)$,
 \ba\label{nx1region3}
&&n_x^{(1)}= -\frac{A_\otimes
\hat{B}_y^{(0)}\psi_0yl}{2\mu_0\hbar\omega_e[1+(z/f)^2]^{1/2}}
   \left\{\frac{k_g}{2(z+f^2/z)}
   \sin{\Phi }\right. \nonumber\\
   &&\qquad\quad\left.+\frac{1}{W_0^2[1+(z/f)^2]}
\cos{ \Phi }\right\} \exp{\left(-\frac{r^2}{W^2}\right)},
\ea
where  $\mu_0$ is the permeability in vacuum,
and
\be
\Phi \equiv
\frac{k_gr^2}{2R}-\arctan{(\frac{z}{f})}.
\ee
The  phase $\delta=\pi/2$ has been taken for concreteness.
As  Eqs.(\ref{nx1region2}) and  (\ref{nx1region3}) show,
 $n_x^{(1)}$ is only produced by the GWs of
$\otimes$-polarization mode,
and is proportional
 to the static magnetic field $\hat{B}_y^{(0)}$
 and  the maximal amplitude   $\psi_0$ of the maser beam.
Since $|n_x^{(1)}|$ contains a decaying factor $e^{-r^2/W^2}$, it
decreases radially for larger $r$. To visualize the dependence of
$n_x^{(1)}$ on spatial variables, we plot it as a function of
$(y,z)$ on the plane $x=0.05$ m in Fig.\ref{nxdependyz}, and  as a
function of $(x,y)$ on the plane $z=0.4$ m in Fig.\ref{nxdependyz1}.

\begin{center}
{ \bf B.   GWs along some other directions }
\end{center}

The above is for the incident GWs along the positive $z$-axis.
 In this section,
 we give  the results of incident
 GWs propagating along  other directions,
while the setup of the detector  is the same as in Section II.

(1) The incident GWs propagating along the negative  $z$-direction.
The relevant results are the following:

  Region II ($0\leq z \leq l$),
\ba\label{nxnegz}
&&n_x^{(1)}=\frac{A_\otimes
\hat{B}_y^{(0)}\psi_0y}{2\mu_0\hbar\omega_e[1+(z/f)^2]^{1/2}}
   \left\{\frac{k_g(z-l)}{2(z+f^2/z)}
   \sin{\left(2k_gz+\Phi\right)}\right.\nonumber\\
&&\qquad\quad+\frac{(z-l)}{W_0^2[1+(z/f)^2]}
\cos{\left(2k_gz+\Phi\right)}\left.+\frac{1}{2(z+f^2/z)}\sin{(k_gz)}
\sin{\left(k_gz+\Phi\right)}\right.\nonumber\\
&&\qquad\quad\left.+\frac{1}{k_gW_0^2[1+(z/f)^2]}\sin{(k_gz)}
\cos{\left(k_gz+\Phi\right)}
    \right\}\exp{\left(-\frac{r^2}{W^2}\right)};
\ea

  Region III ($z>l$),
\ba
 n_x^{(1)}=0.
\ea
Since the maser emitter lies in  Region I ($z<0$), so it is not
interesting for detection.

(2) The  incident  GWs propagating  along
 the positive  $x$-direction.
 The  static magnetic field
 $\hat{B}^{(0)}_y$  is taken to
 be localized in the region $-l_1\leq x \leq l_2$
 in the $x$-direction.
One obtains
 \ba\label{nxx}
&&n_x^{(1)}=\frac{A_\otimes
\hat{B}_y^{(0)}\psi_0y}{2\mu_0\hbar\omega_e[1+(z/f)^2]^{1/2}}
   \left\{\frac{k_g(x+{l_1})}{2(z+f^2/z)}
   \sin{[k_g(z-x)+\Phi]}\right.\nonumber\\
&&\qquad\quad+\frac{x+l_1}{W_0^2[1+(z/f)^2]}
\cos{[k_g(z-x)+\Phi]}\left.+\frac{1}{2(z+f^2/z)}\sin{(k_gx)}
\sin{\left(k_gz+\Phi\right)}\right.\nonumber\\
&&\qquad\quad\left.+\frac{1}{k_gW_0^2[1+(z/f)^2]}\sin{(k_gx)}
\cos{\left(k_gz+\Phi\right)}
    \right\}\exp{\left(-\frac{r^2}{W^2}\right)}
 \ea
for the region $-l_1\leq x \leq l_2$.

(3) The  incident  GWs propagating  along the negative
$x$-direction. With the static magnetic field as in (2), one has
 \ba\label{nxx1}
  &&n_x^{(1)}=-\frac{A_\otimes
\hat{B}_y^{(0)}\psi_0y}{2\mu_0\hbar\omega_e[1+(z/f)^2]^{1/2}}
   \left\{\frac{k_g(x-l_2)}{2(z+f^2/z)}
   \sin{[k_g(z+x)+\Phi]}\right.\nonumber\\
&&\qquad\quad+\frac{x-l_2}{W_0^2[1+(z/f)^2]}
\cos{[k_g(z+x)+\Phi]}\left.+\frac{1}{2(z+f^2/z)}\sin{(k_gx)}
\sin{\left(k_gz+\Phi\right)}\right.\nonumber\\
&&\qquad\quad\left.+\frac{1}{k_gW_0^2[1+(z/f)^2]}\sin{(k_gx)}
\cos{\left(k_gz+\Phi\right)}
    \right\}\exp{\left(-\frac{r^2}{W^2}\right)},
 \ea
which is  similar to Eq.(\ref{nxx}).

(4) The  incident  GWs propagating  along  the positive or
   negative $y$-direction.
One finds  that $\tilde{E}_y^{(1)}= 0$, leading to
\be
 n_x^{(1)}=0.
\ee

The above results show that, for the given setup,
the detector
responses differently to the incident GWs  from  different directions.
In the following it will be seen that
the detector responses most effectively to the GWs in the $z$-direction.
In general, RGWs is of stochastic nature and
come from various directions.
In this case,
a reduction factor will be introduced as shown later.

\begin{center}
{ \bf C.   Numerical calculations for PPFs}
\end{center}

In order to examine the dependence of perturbed photons on the directions
in which  GWs propagate,
let us estimate numerically the  perturbed photons per unit time
received by the microwave receiver
for the above cases of the incident GWs.
For concreteness,
we adopt the following parameters of the detector that can be
realized in the laboratory:

1) $P=2$\,kW, the power of the maser beam, corresponding to
$\psi_0=1.8\times 10^4$ Vm$^{-1}$ for the spot radius $W_0=0.05\,$m
\cite{Cohen,Abramovich}.

2) $\hat{B}_y^{(0)}=3$\,Tesla, the strength of the background
static magnetic field \cite{Perenboom}.

3) $l =0.4\,$m, the width of the static magnetic field in
$z$-direction.

4) $l'=l_1+l_2 =0.4\,$m, the width of the static magnetic field in
$x$-direction.

5) $\nu_e \simeq 4.5$\,GHz, the frequency of the maser beam
  in the microwave band \cite{Cohen,Abramovich}.

The number of perturbed photons in the $x$-direction per second
passing through a  surface
$\Delta s$ on the plane $x=0.05$m is given by:
 \be\label{total}
 N_x^{(1)}=\int\limits_{\Delta s}n_x^{(1)}|_{x=0.05}\
{\d}y{\d}z.
 \ee
Here the integrand is taken to be
the negative portion of $n_x^{(1)}<0$,
which is reflected by the membrane back to the positive $x$-direction.
For comparison, we  choose
$\Delta s   \simeq 8\times10^{-2}\,$m$^{2}$ ($0\leq y \leq 0.2\,$m,
$0\leq z\leq 0.4\,$m)
to receive more photons with a limited size.
For concreteness, $A_\otimes\sim10^{-30}$ is taken.
The resulting  $N_x^{(1)}$ is shown in Table 1.
We see that the
 magnitude of $N_x^{(1)}$ generated
 by the incident GWs along the positive $z$-direction and
 the positive $x$-direction has the same order,
 which is  larger than that for other cases.
Note, for the case of GWs  along
$y$-direction, $N_x^{(1)}$  is nearly vanishing.

Using the parameters given above, if we choose the
maser beam to have a polarization degree is $\sim 98\%$,
corresponding to a ratio of unpolarized/polarized electric field
components $\tilde{E}_y^{(0)}/\tilde{E}_x^{(0)}\sim 0.1$,
 our computation shows that the ratio
of the number of background/perturbed photons  over the area $\Delta
s$ per second will be \be N_x^{(0)}/N_x^{(1)}\sim 10^{-7} \ee at
$x\simeq 1\,m$ . So, if  the microwave receiver is put at $\sim 1$
meter away from the fractal membrane, the influence of the
background photon flux will be effectively negligible. A higher
polarization of the  maser beam is always wanted to suppress the
BPF.

\begin{table}
 \caption{\label{table1}
The number of photons per second passing through $\Delta s$,  $N_x^{(1)}$,
 for GWs from various directions.}
 \begin{center}
\begin{tabular}{c|c}
 \hline
 Direction of GWs & $N_x^{(1)} (s^{-1})$ \\
 \hline
 $+z$ & $ \sim1\times10^2$ \\
 $-z$&$ \sim5.4\times10$\\
 $+x$&$  \sim1.2\times10^2$\\
 $-x$&$  \sim3.4$\\
 $\pm y$&   $\sim  0$\\
 \hline
\end{tabular}
\end{center}
\end{table}

\begin{figure}
\centerline{\scalebox{1.6}[1.6]
{\includegraphics[width=8cm]{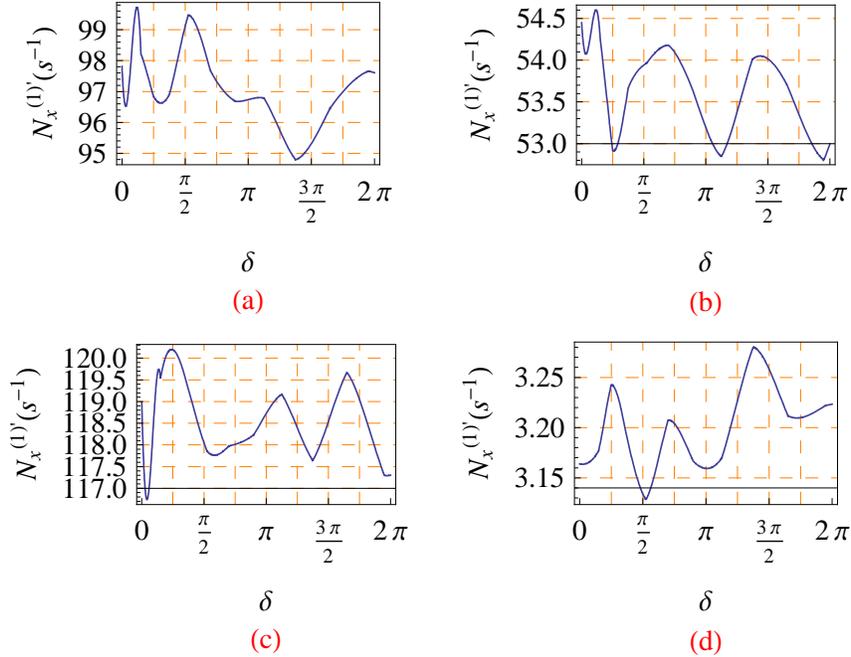}}} \caption{  \label{phase}
The small fluctuation of  $N_x^{(1)}$  with variable $\delta$.
(a),(b),(c) and (d) give the cases of the incident
GWs along $+z, -z, +x$ and $-x$, respectively.   }
\end{figure}

Remember that the phase  $\delta=\pi/2$ has been taken in the  above
for simplicity.
However, in general, $N_x^{(1)}$ would
depend on the phase factor $\delta$. Fig.{\ref{phase}} (a),(b),(c)
and (d) give $N_x^{(1)}$ as a function of $\delta$ for the incident
GWs along $+z, -z, +x$ and $-x$, respectively.
Fig.{\ref{phase}} shows that
the changes of $N_x^{(1)}$ with $\delta$ is small,
and the error for various values of  $\delta$ is less than $4\%$.
Thus, in the following,
we assume that $N_x^{(1)}$ is independent of the phase factor $\delta$.

\begin{center}
{\bf IV.   THE  DETECTION   FOR  RGWs }
\end{center}

In this section, we estimate the sensitivity of the detector, and
analyze the feasibility of detecting the RGWs using this method.

\begin{center}
{ \bf A.  Implementation of the experiment}
\end{center}

One certainly expects to have a number of problems in the actual
implementation of such kind of detection.
There could be various sorts of noises for this detector,
such as  thermal noise,  external EM  noise,
seismic noise, and  shot noise in the maser beam, etc.
The seismic noise is one of major obstacles for
the common laser interferometer detectors.
However, this kind of noise usually
has a  frequency much lower than GHz band
and will not generate additional perturbed photons,
so it will not affect our detection essentially.
Still,  an isolating system may be employed.
For instance, the detector system may be put on
a suspended framework to absorb seismic vibrations.
As for the shot noise in the maser beam,
it can be suppressed by stabilizing the frequency
and the amplitude of the maser beam.

Among all these kinds of noises, the external EM noise would be a
real problem for our detection. For example, the CMB at $T\sim 2.7$
K yields a photon flux density $n\sim 10^{4}$ $cm^{-2} s^{-1}$
around the frequency $\nu \sim 4.5$ GHz with a width $\Delta \nu
\simeq 10$ KHz (the frequency width of the maser) in any direction.
The energy flux  of CMB photons  would overwhelm that of the
perturbed photons, since the average PPF density $\bar{n}_x^{(1)}
\sim 0.6$ $cm^{-2} s^{-1}$ as from Table \ref{table1} where we have
assumed  $A_\otimes\sim10^{-30}$.
To solve the problem, we propose
to employ a Faraday cage, which shields the detecting device from
the external EM noise. The outer shell of the cage may be made of
some conducting metal so that the external EM waves will not
entering the cage.
The inner surface of the cage should be made of
some kind of material that effectively absorbs noise photons within
the cage. So the CMB photons and other EM noise inside the cage can
be eliminated.
In designing the Faraday cage, we should excavate a
little hole on the cage and the hole should be sealed by a one-way
membrane with a total transmittance around $4.5$ GHz, so that the
maser beam can pass through it while the photons can not enter the
cage.

Furthermore, to eliminate thermal photons emitted from the detector
system and from the inner layer of the cage, one need reduce the
temperature of the system. Therefore, a cryogenic technique should
be applied so that the detector operates in a low temperature
environment. Moreover, a vacuum environment of the system will help
for the detection.

\begin{center}
{ \bf B.  Sensitivity of detector }
\end{center}

 For a preliminary analysis, we will focus on the thermal
noise in our detection system and estimate
its sensitivity limited by  thermal noise.
 The signal power  is given by
 \be
 S=\eta  |N_x^{(1)}|\hbar\omega_e,
 \ee
where $\eta$ is the reflectance of the fractal
membranes, ranging from $0$ to $1$.
With the help of  Eq.(\ref{total}),
$N_x^{(1)}$ is given by
\be \label{photon}
 |N_x^{(1)}|\simeq\frac{7A_\otimes \hat{B}^{(0)}_y\psi_0}
  {\mu_0\hbar\omega_e}\times10^{-3} \,\,\, s^{-1}.
 \ee
The input thermal noise (the thermal noise
coming into the input part of the receiving system)
can be estimated
as
 \be
N_{\rm{in}}=k_BT B,
 \ee
where $k_B$ is Boltzmann's constant, $T$ is the temperature of the
thermal noise, and  $B$ is the bandwidth of the detector in Hz,
which can be estimated as $B\sim \nu_e/Q$, where $Q$ is its
quality factor.
There are  additional thermal noises within  the
receiving   system, thus the minimal signal power should satisfy
\cite{Skolnik}
 \be\label{ratio}
 S_{\rm min} =M f_0 N_{\rm{in}},
 \ee
where $M\geq1$ is the minimal output signal-to-noise,
and $f_0>1$ is the noise coefficient of the microwave receiver,
defined as the
ratio of the input signal-to-noise to the output signal-to-noise.
Then using Eqs.(\ref{photon})-(\ref{ratio}) and letting $M=1$
yield the minimal detectable dimensionless amplitude,
\be   \label{min}
h_{\rm{min}}\simeq\frac{6.4\mu_0k_BT f_0}
     {\eta\hat{B}^{(0)}_y\psi_0Q}\times10^{11},
\ee where $h_{\rm{min}}\equiv A_{\otimes\, \rm{min}}$. Taking
$\psi_0=1.8\times 10^4$ Vm$^{-1}$ and  $\hat{B}_y^{(0)}=3$ Tesla,
$Q=10^5$, $T\sim 1$mK \cite{Zemansky}, $f_0\sim 2$, $\eta \sim
99.99\%$ \cite{Wen,Zhou,Hou}, one obtains the sensitivity \be
\label{hmin} h_{\rm{min}}\simeq  4 \times10^{-30}. \ee

As said earlier,
RGWs come from all directions and form a
stochastic background, therefore,
in evaluating  the sensitivity of the detector,
a reduction factor $F$ should be introduced \cite{Maggiore}.
We can estimate its magnitude as follows.
Firstly,  we consider the case that the detector responses only to
the incident GWs along the positive $z$-direction.
By Eqs.(\ref{firstorder4}) and (\ref{nxxx}) in Appendix,
\be
n_x^{(1)}\propto
\tilde{E}_{\hat{y}}^{(1)} \propto  \partial h_\otimes(z,t)/\partial
z \, \propto k_z h_\otimes(z,t).
\ee
For a beam of incident GWs with a wave vector  $\bf k$,
one needs to project it along $z$-direction,
so that its component
$n_x^{(1)}\propto k_z = k \cos \theta $,
where $\theta$ is the angle between $\bf k$
and the positive $z$-direction.
Then in this special case
the reduction factor $F$ will be estimated as
\be
F=\frac{1}{4\pi}
  \int_0^{\frac{\pi}{2}}  \cos{\theta}
   \sin\theta d\theta \int_0^{2\pi} d\phi  = \frac{1}{4}.
\ee
However, this estimate is not complete.
In fact, while the detector does not response to
the incident RGWs in the  $y$-direction,
it responses to
the incident RGWs in both the  $x$- and $z$-directions
with the same order of magnitude,
as shown in  Table \ref{table1}.
For an qualitative estimation,
we can take the response of the detector
to any incident GWs perpendicular to the $y$-direction
to be the same.
Consider  an arbitrary  beam of GWs whose wave vector $\bf k$
forms an angle $\theta $ with the $y$-direction.
Projecting the wave vector $\bf k$ on the $x-z$ plane
gives rise to a factor $\sim k \sin \theta$.
Then, taking the average of $\sin \theta$ over the solid angle $4\pi$
yields
\be \label{reduction}
F =\frac{1}{4\pi}\int_0^\pi \sin^2{\theta} d\theta
\int_0^{2\pi} d\phi = \frac{\pi}{4}.
\ee
Therefore, we expect that the actual reduction factor $F$ for our detector
would be between $1/4$ and $\pi/4$. Multiplying Eq.(\ref{photon})  by  $F$,
 the sensitivity of the detector given by Eq.(\ref{hmin})
should be modified  as
\be
\label{hmin1}
h_{\rm{min}}\simeq  (5.1 \times10^{-30}\sim 1.6\times 10^{-29}).
\ee

\begin{center}
{ \bf C.  Detecting  RGWs}
\end{center}

\begin{figure}
\centerline{\scalebox{1.0}[1.0]
{\includegraphics[width=10cm]{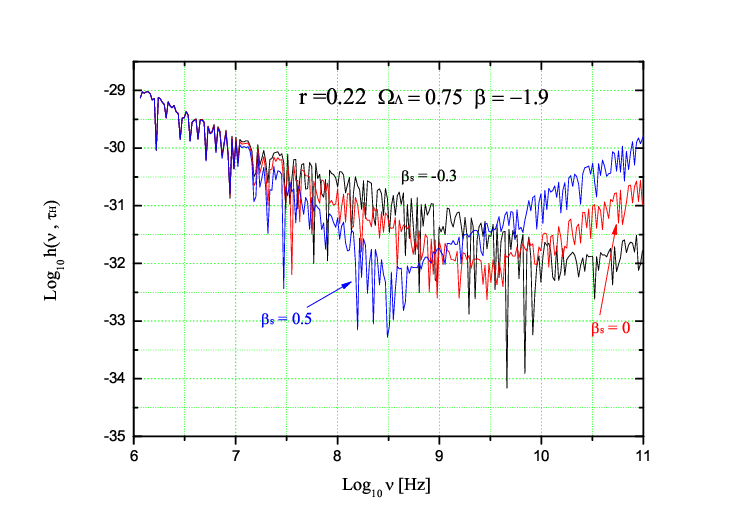}}} \caption{\label{1.8} The
spectrum of RGWs for the cosmological model with the tensor-scalar
ratio $r=0.22$,
 the dark energy $\Omega_\Lambda=0.75$, and the inflation
parameter $\beta=-1.9$.
 The spectrum in  GHz band
depends sensitively on the reheating parameter $\beta_s$
\cite{Miao}.  }
\end{figure}

What about the detection target, say,
the the RGWs in the present accelerating
universe \cite{Zhang2,Zhang1,Zhang3} around the frequency
$\nu_g\sim4.5$ GHz?
Now we  calculate the root-mean-square (r.m.s.) amplitude of RGWs.
In the high frequency limit,
the RGWs can be considered  approximately
as the superposition of plane waves in Eq.(\ref{circular}).
By its nature,  RGWs constitute a stochastic background,
and the mean value of the field $h_{ij}$ is
zero at every instance of time and at every spatial point:
$ \langle0|h_{ij}({\x},\tau)|0 \rangle=0$.
But the variance is not zero \cite{Grishchuk1,Zhang2,Zhang1},
\be\label{hms}
 \langle h^2\rangle\equiv\langle0|h^{ij}({\x},\tau)\,
h_{ij}({\x},\tau)|0\rangle
\equiv
\int_0^\infty
h^2(\nu,\tau)\frac{{\d}\nu}{\nu},
\ee
where $h(\nu,\tau)$ is the spectrum of the RGWs,
and $\tau$ is the conformal time.
Fig.\ref{1.8} gives the spectrum $h(\nu,\tau_H)$
of the RGWs at the present time $\tau_H$
for the cosmological model with the tensor-scalar ratio $r=0.22$,
the dark energy $\Omega_\Lambda=0.75$,
the inflation parameter $\beta=-1.9$,
and the reheating parameter $\beta_s$ \cite{Miao}.
The quantity $h(\nu,\tau_H)$ is
related to  the spectral energy density $\Omega_g(\nu)$
 often used in literatures\cite{Grishchuk1,Miao,Maggiore},
\be\label{density}
\Omega_g(\nu)=\frac{\pi^2}{3}(\frac{\nu}{\nu_H})^2h^2(\nu,\tau_H),
\ee
where $\nu_H=H_0\sim 2\times 10^{-18}$ Hz is the Hubble frequency.

Due to the resonance condition,
the detector only responses to a very narrow frequency band
 $\Delta\nu \simeq \nu_g/Q$ around the central frequency $\nu_g $,
 where $Q$ is the quality factor of the maser beam.
Thus, only the modes of frequencies $\nu_g\simeq 4.5$ GHz
are selected among the incident RGWs.
The integration in Eq.(\ref{hms}) is then evaluated as
\be
\int_0^\infty h^2(\nu,\tau_H)\frac{{\d}\nu}{\nu} \simeq
h^2(\nu_g,\tau_H) \frac{\Delta \nu}{\nu_g} \simeq
h^2(\nu_g,\tau_H)/Q.
\ee
So the r.m.s. amplitude of the RGWs in the band is
\be  \label{hms1}
h_{\rm rms} \equiv \sqrt{\frac{\langle h^2\rangle}{2}}\simeq
\frac{h(\nu_g,\tau_H )}{\sqrt{2Q}},
\ee
where $\sqrt{2}$ accounts for the assumption that
the $\otimes$- and  $\oplus$-polarization modes
give equal contribution.
Reading from the known spectrum in Fig.\ref{1.8} gives
\be \label{infl}
h(\nu_g,\tau_H)\sim8\times10^{-32}
\ee
at $\nu\sim 4.5$ GHz  for the reheating model  $\beta_s=0.5$.
The corresponding r.m.s. amplitude is
\be\label{hfact}
h_{\rm rms}\simeq1.8\times10^{-34}.
\ee
Comparing Eq.(\ref{hmin1}) with Eq.(\ref{hfact}),
one can see that there is approximately
a gap of $4\sim 5 $ orders of magnitude between
the sensitivity of the sample detector and
the  r.m.s.amplitude of RGWs in the accelerating universe.

As for the possible thermal background of gravitational waves
\cite{Zeldovich,buonanno},
it could have been generated at the
very early stage at an energy scale $\sim 10^{19}$ Gev, also
described by the Planck spectrum like CMB photons. If there is no
inflationary process, the graviton gas would be at $T\sim 1 $  K,
corresponding to typical  frequencies  in $\sim 100$ GHz.
The amplitude of the spectrum of this thermal GWs would be  about
$h(\nu) \sim 10^{-32}$ around  the frequency of  $4.5$ GHz.
However, if the inflationary expansion has occurred by
some $60$ e-folding, as is supported by WMAP data of CMB
anisotropies \cite{spergel} and others,
the thermal GWs would be drastically diluted and
its  temperature would be reduced  to $T\sim 10^{-28}$ K,
totally negligible.

\begin{center}
{ \bf D.  Possible improvements }
\end{center}

Although the sensitivity is still $4\sim5$ orders short to detect
the  RGWs, the detector has a large room for improve in  several
ways.
Firstly,  note that $h_{\rm min}\propto 1/Q$ by
Eq.(\ref{min}), while $h_{\rm rms}\propto  1/\sqrt{Q}$ by
Eq.(\ref{hms1}), so the ratio $h_{\rm rms}/ h_{\rm min} \propto
\sqrt{Q}$. A larger quality factor $Q$ of maser beam will enhance
the possibility for detection. This would require a highly
monochromatic maser beam. For instance, if $Q$ can be increased from
$10^5$ to $10^{9}$, the ratio would be enhanced by $\sim 100$ times
and the gap will be reduced by $2$ orders. At present, for
conventional lasers in the optical frequency band, a quality factor
$Q\sim10^{13}$  has been achieved \cite{Barber}, and for hydrogen
maser, the  quality factor $Q$ has been reached up to $\sim10^9$
\cite{Harry,David,Kleppner}. Secondly, as can be seen from
Eq.(\ref{min}), the sensitivity depends strongly on the temperature
$T$ of detector. If it is reduced down to $T\sim 50\mu$K
\cite{Lounasmaa}, the sensitivity will be  improved by a factor
$\sim 20$ and the gap will be suppressed to be about 1 order.
Thirdly, increasing the strength of the static magnetic field
$\hat{B}_y^{(0)}$ and the power of the maser will also improve the
sensitivity of the detector. Apart from the above possible
improvements, enlarging the interaction dimension between the GWs
and the static magnetic field will also improve the detection
\cite{li2}. Putting these possible improvements together,  an actual
detection will be  realistic.

\begin{center}
{\bf V. CONCLUSION  AND  DISCUSSION}
\end{center}

We have extensively studied the maser beam method for detection of
GWs $\sim 4.5$ GHz.
The experimental setup consists of a maser beam,
a strong static magnetic field,
a reflecting fractal membrane and a microwave receiver.
Moreover, a Faraday cage should be used
to prevent the detector from  external EM noises.
And, to reduce thermal noise,
the detector should be place in a low temperature environment.
The maser beam is chosen to be linearly polarized,
so that the BPF in the detecting direction can be suppressed effectively,
and the PPF can be detected as a signal of GWs.
We have obtained the analytical expressions for the PPF density $n_x^{(1)}$
generated by the incident GWs from various directions.


To examine the feasibility of the detection, we have estimated the
sensitivity of the sample  detector limited by thermal noise
 and have confronted it with the
RGWs  in the accelerating universe as a scientific object.
In our preliminary analysis, we found that there was still
 a gap of about $4\sim 5$ orders between the  sensitivity of
the detector and the r.m.s. amplitude of the RGWs. However, we have
a lot of ways in improving
 the sensitivity, such as  lowering the  temperature, increasing
the quality factor and the power of maser beam, and enlarging the
strength and dimension of the static magnetic field.
These improvements will remove the gap, making the  method applicable for
detecting high frequency RGWs.

However, our analysis on the
detection for the RGWs are still tentative, and the conclusions
arrived are also preliminary. In particular, the odds is that the
detecting PPF as the signal is very small and confronts  a number of possible
sources of EM noise. A systematical analysis on effectively
suppressing these noises is then needed to give a more reliable
sensitivity.

Overall,
the maser beam method in GHz band or higher is feasible.
As a new method to detect GWs,
it is complementary to the laser interferometer method
 working in the low frequency range $(10^{-4}\sim 10^4)$ Hz.
Moreover, from the
point of view of experimental constructions,
the building of this
detection is much less expensive than ordinary interferometer laser
methods. Therefore, under these considerations,
the GW detection scheme is certainly worthy of further
studies and is expected to be implemented in laboratory  someday.

\begin{center}
{\bf ACKNOWLEDGMENTS}
\end{center}

Y. Zhang's work was supported by the CNSF
No.10773009, SRFDP, and CAS.
F.Y. Li's work was sopported by
CNSF No.10575140 and National Basic Research Program of China
under Grant No.2003B716300.

\begin{center}
{\bf APPENDIX: CALCULATIONS OF $n_x^{(1)}$ FOR GWs ALONG DIRECTION
OF MASER BEAM}
\end{center}

In this appendix, we present the calculations of the PPFs density
$n_x^{(1)}$ produced by the GWs along the positive $z$-direction. By
Eq. (\ref{initial2}), the space is divided into three regions:
 I $(z< 0)$,
  II $(0\leq z\leq l)$,
   and III $(z> l)$.
Firstly, we  focus on the region II,
 where the static magnetic
field $\hat{B}_y^{(0)} \ne 0$. By Eq.(\ref{circular}), the
expressions of GWs only have two variables $(z,t)$, so will  be the
perturbed EM fields accordingly. Plugging  Eqs. (\ref{metric}) -
(\ref{fcovariant2})
into Eqs. (\ref{Mexwell}) and (\ref{Mexwell2}),
and
keeping only up to the linear  terms of $h_{\mu\nu}$, then after
lengthy but easy calculations,  one obtains
\ba\label{firstorder1}
 &&\frac{1}{c^2}\tilde{E}^{(1)}_{x,\,t}+\tilde{B}^{(1)}_{y,\,z}
 =\hat{B}^{(0)}_yh_{\oplus,\,z}\\
 &&\tilde{E}^{(1)}_{x,\,z}+\tilde{B}^{(1)}_{y,\,t}=0\\ \label{firstorder4}
 &&\frac{1}{c^2}\tilde{E}^{(1)}_{y,\,t}-\tilde{B}^{(1)}_{x,\,z}
 =\hat{B}^{(0)}_yh_{\otimes,\,z}\\ \label{firstorder2}
 &&\tilde{E}^{(1)}_{y,\,z}-\tilde{B}^{(1)}_{x,\,t}=0,
 \ea
 and
 \be\label{z}
 \tilde{E}^{(1)}_{z,\,t}=\tilde{E}^{(1)}_{z,\,z}=0,\quad
  \tilde{B}^{(1)}_{z,\,t}=\tilde{B}^{(1)}_{z,\,z}=0.
 \ee
Using Eq.(\ref{circular}),
one solves Eqs.(\ref{firstorder1})-(\ref{firstorder2})
in Region II \cite{li1}:
 \ba\label{ex}
&&\tilde{E}_x^{(1)}=
 \frac{i}{2}A_\oplus \hat{B}_y^{(0)}k_gc z
e^{i(k_gz-\omega_gt)} +b_1 e^{i(k_gz-\omega_gt)}
+c_1 e^{i(k_gz+\omega_gt)},\nonumber\\
&& \tilde{B}_y^{(1)}= \frac{i}{2}A_\oplus \hat{B}_y^{(0)}k_g z
e^{i(k_gz-\omega_gt)} + b_2 e^{i(k_gz-\omega_gt)}
 +c_2 e^{i(k_gz+\omega_gt)},\\
&&\tilde{E}_y^{(1)}=-\frac{1}{2}A_\otimes \label{ey}
\hat{B}_y^{(0)}c k_g z e^{i(k_gz-\omega_gt)}
+ib_3 e^{i(k_gz-\omega_gt)}+ic_3 e^{i(k_gz+\omega_gt)},\nonumber\\
&&\tilde{B}_x^{(1)}=\frac{1}{2}A_\otimes \hat{B}_y^{(0)}k_g z
e^{i(k_gz-\omega_gt)} +ib_4 e^{i(k_gz-\omega_gt)}+ic_4
e^{i(k_gz+\omega_gt)}.
 \ea
From Eq.(\ref{z}), one obtains  a physical solution,
\be \label{ez}
 \tilde{E}_{z}^{(1)} =\tilde{B}_{z}^{(1)}=0,
\ee
which is valid in  all the  three regions.
The constants, $b_1$, $c_1$, ... $b_4$, $c_4$,
in Eqs.(\ref{ex}) and (\ref{ey})
are to be determined  by
 the physical requirements and boundary conditions
 in the following.

Any physical measurement by an observer in curved spacetime should
be carried out in a local inertial frame, i.e., the observable
quantities are the projections of the physical quantities
 on to the four orthonormal bases
$e_{\hat{0}}^\mu,e_{\hat{1}}^\mu,e_{\hat{2}}^\mu,e_{\hat{3}}^\mu$
carried by the observer.
Therefore, the observable  EM fields are
\be
\label{measure}
F_{\hat{\alpha}\hat{\beta}}=F_{{\mu}{\nu}}e_{\hat{\alpha}}^\mu
e_{\hat{\beta}}^{\nu}.
\ee
For an observer at rest with respect to
the static magnetic field,
one can
choose \ba\label{space}
 && e_{\hat{0}}^\mu=(1,0,0,0), \nonumber \\
 &&e_{\hat{1}}^\mu=(0,1-\frac{1}{2}h_{\oplus},0,0),\nonumber \\
 &&e_{\hat{2}}^\mu=(0,-h_{\otimes},1
      +\frac{1}{2}h_{\oplus},0),\nonumber \\
 &&e_{\hat{3}}^\mu=(0,0,0,1).
\ea
Suppose that, for simplicity,
there is no  perturbed EM waves propagating in the negative $z$ direction
in Region I and Region III
 \cite{li1}.
From Eqs.(\ref{fcovariant1}),
(\ref{measure})  and (\ref{space}),
and by the boundary conditions
that the real parts of the perturbed fields $F_{\mu\nu}^{(1)}$
are continuous at the interfaces,
the observable perturbed EM fields
in  the three regions are given by:

  Region I $(\hat{B}^{(0)}=0)$:
  \ba
\tilde{E}_{\hat{x}}^{(1)} =\tilde{E}_{\hat{y}}^{(1)} =
\tilde{B}_{\hat{x}}^{(1)} =\tilde{B}_{\hat{y}}^{(1)}=0;
  \ea

  Region II $(\hat{B}^{(0)}=\hat{B}_y^{(0)})$:
 \ba\label{ExBy12}
&&\tilde{E}_{\hat{x}}^{(1)}=\frac{i}{2}A_\oplus
\hat{B}_y^{(0)}ck_g z e^{i(k_gz-\omega_gt)},\nonumber\\
&& \tilde{B}_{\hat{y}}^{(1)}=\frac{i}{2}A_\oplus \hat{B}_y^{(0)}k_gz
e^{i(k_gz-\omega_gt)},\\ \label{EyBx12}
&&\tilde{E}_{\hat{y}}^{(1)}=-\frac{1}{2}A_\otimes
\hat{B}_y^{(0)}ck_g z e^{i(k_gz-\omega_gt)}\nonumber\\
&&\qquad\quad+\frac{i}{4}A_\otimes \hat{B}_y^{(0)}c
e^{i(k_gz-\omega_gt)}+\frac{i}{4}A_\otimes
\hat{B}_y^{(0)}c e^{i(k_gz+\omega_gt)},\nonumber\\
&&\tilde{B}_{\hat{x}}^{(1)}=\frac{1}{2}A_\otimes \hat{B}_y^{(0)}k_gz
e^{i(k_gz-\omega_gt)}\nonumber\\
&&\qquad\quad+\frac{i}{4}A_\otimes \hat{B}_y^{(0)}
e^{i(k_gz-\omega_gt)}+\frac{i}{4}A_\otimes \hat{B}_y^{(0)}
e^{i(k_gz+\omega_gt)}.
 \ea
Note that the expressions of $\tilde{E}_{\hat{x}}^{(1)}$ and
$ \tilde{B}_{\hat{y}}^{(1)}$ in Eq.(\ref{ExBy12}) are different
from those given by Eq.(43) in Ref.\cite{li1}.

 Region III $(\hat{B}^{(0)}=0)$:
 \ba\label{ey1}
&&\tilde{E}_{\hat{x}}^{(1)}=\frac{i}{2}A_\oplus
\hat{B}_y^{(0)}c k_g l e^{i(k_gz-\omega_gt)},\nonumber\\
&&\tilde{B}_{\hat{y}}^{(1)}=\frac{i}{2}A_\oplus
\hat{B}_y^{(0)}k_g l e^{i(k_gz-\omega_gt)},\\ \label{ey1}
&&\tilde{E}_{\hat{y}}^{(1)}=-\frac{1}{2}A_\otimes
\hat{B}_y^{(0)}c k_g l e^{i(k_gz-\omega_gt)},\nonumber\\
&&\tilde{B}_{\hat{x}}^{(1)}=\frac{1}{2}A_\otimes
\hat{B}_y^{(0)}k_g l e^{i(k_gz-\omega_gt)},
 \ea
 where $l$ satisfies
 \be
l=n\lambda_g \qquad (n \    {\mathrm{is \ an \ integer}}).
 \ee
 It is straightforward to obtain from Eq.(\ref{ez})
 \be
\tilde{E}_{\hat{z}}^{(1)} =\tilde{B}_{\hat{z}}^{(1)}=0,
 \ee
also valid in all the three regions. From the perturbed EM fields
given above, it is straight forward to obtain the PPFs density:
 \be\label{nxxx}
n_x^{(1)}= \frac{1}{\mu_0\hbar\omega_e}\langle
\tilde{E}_{\hat{y}}^{(1)}\tilde{B}_{\hat{z}}^{(0)} \rangle_{
\nu_g=\nu_e},
 \ee
where the subindex ``$\nu_g=\nu_e$''
indicates  the resonance  condition,
under which the time average will be non-vanishing.

\end{document}